\documentclass[pdflatex,sn-mathphys-num]{sn-jnl}

\usepackage{graphicx}%
\usepackage{multirow}%
\usepackage{amsmath,amssymb,amsfonts}%
\usepackage{amsthm}%
\usepackage[title]{appendix}%
\usepackage{xcolor}%
\usepackage{textcomp}%
\usepackage{manyfoot}%
\usepackage{booktabs}%
\usepackage{algorithm}%
\usepackage{algorithmicx}%
\usepackage{algpseudocode}%
\usepackage{listings}%
\usepackage{lmodern}
\usepackage{bm}

\begin{document}

\title[Title]{Fast Prediction of Phase shifters Phase Error Distributions Using Gaussian Processes}


\author*[1]{\fnm{Zijian} \sur{Zhang}}\email{hectorzhang4253@gmail.com}

\affil*[1]{\orgname{University of Electronic Science and Technology of China}, 
\orgaddress{\city{Chengdu}, \postcode{611731}, \country{China}}}


\abstract{We present an efficient method for evaluating random phase errors in phase shifters within photonic integrated circuits, avoiding the computational cost of traditional Monte Carlo simulations.   By modeling spatially correlated manufacturing variations as a stationary Gaussian process and applying linear functional theory, our approach enables rapid prediction of phase difference distributions through numerical integration.   We validate the method using tolerance optimization and virtual manufacturing experiments, demonstrating its accuracy and significant reduction in computational burden.   This framework offers a practical and scalable alternative for error analysis and robust design in photonic systems.}




\maketitle
\newpage
\section{Introduction}\label{sec1}

Phase shifters are key components in photonic integrated circuits (PICs), where manufacturing imperfections often induce variations in waveguide geometry and refractive index, leading to unintended phase shifts or delays that degrade device performance.

One representative example is the Mach-Zehnder interferometer (MZI), where fabrication-induced waveguide width variations between the two arms lead to phase errors that can shift the operating wavelength and impact interference-based functionality. On thin-film lithium niobate platforms, such phase errors can often be calibrated via DC bias tuning. However, silicon photonic platforms primarily rely on thermo-optic phase shifters for post-fabrication calibration. While effective, thermal tuning adds significant system complexity and static power consumption, which becomes increasingly problematic as system scale grows. In large MZI or MRR arrays, accumulated random phase errors pose serious challenges to system control and calibration\cite{gao_automatic_2020,khan_optimized_2022}.

To improve tolerance and reduce the need for thermal tuning, various design strategies have been explored—such as using wider waveguides \cite{song_machzehnder_2021,song_toward_2022} and exploiting spatial correlations via folded MZI structures\cite{bogaerts_predicting_2019}. However, accurately evaluating the statistical phase error distribution in such systems remains computationally challenging. Traditional Monte Carlo simulations can model spatially correlated fabrication errors\cite{bogaerts_layout-aware_2019} and provide accurate estimates of phase distributions with low computational efficiency relatively . Existing non-Monte Carlo methods\cite{murao_compact_2022,voronkov_design_2022,xing_capturing_2022}, while computationally efficient, often neglect spatial correlations in fabrication variations—failing to capture the fact that physically adjacent structures tend to experience similar deviations.

To address these limitations, we propose a fast and accurate method based on stationary Gaussian processes for modeling fabrication-induced phase errors in arbitrary phase-shifter configurations. We model the fabrication variation as a spatially correlated Gaussian process characterized by a covariance function that captures the decay of correlation over distance. The phase response of each component—defined as a weighted path integral over the spatially varying process—is thus a linear functional of the Gaussian field. This allows the resulting phase shift vector to be modeled as a multivariate Gaussian, whose mean and covariance matrix can be computed analytically using the covariance kernel and the spatial support of each path. By further processing the mean and covariance matrix, we can effectively obtain the complete distribution of the device performance such as random phase difference error and random phase gradient error.

Our results show that this method not only preserves the spatial correlation structure inherent to the fabrication process but also significantly reduces computational costs, offering a powerful tool for the robust design and fast tolerance analysis of large-scale photonic systems.
\section{Random Phase Error Theory}\label{sec2}

"Random Phase Error" refers to the additional phase deviation relative to the designed nominal phase, caused by random factors such as manufacturing errors.

Manufacturing errors first alter the effective refractive index $n_{eff}$, which in turn affects the propagation constant $\beta$. The deviation in the propagation constant accumulates along the propagation direction during transmission, ultimately resulting in random phase errors. To facilitate analysis and research, we first define the sensitivity of the effective refractive index as:
\begin{equation}
 {{\xi }_{w}}=\frac{\partial {{n}_{eff}}}{\partial w},{{\xi }_{h}}=\frac{\partial {{n}_{eff}}}{\partial h}\ 
 \label{def:xi}
\end{equation}

Where $\lambda$ is the wavelength, w and h are the width and thickness parameters of the waveguide respectively, and $n_{eff}$ represents the effective refractive index. The random phase error of one arm of MZI can be calculated using the following formula:
\begin{equation}
{{\phi }_{r}}={{k}_{0}}\int_{C}{\left( {{\xi }_{w}}\left( \mathbf{r} \right){{\delta }_{w}\left( \mathbf{r} \right)}+\xi_{h}\left( \mathbf{r} \right) {{\delta }_{h}}\left( \mathbf{r} \right) \right)ds}
 \label{def:phi_r}
\end{equation}
Where $k_{0}$ is the vacuum wave number, sensitivity $\xi_{w}$ and $\xi_{h}$ are related to the waveguide width at coordinates x,y, $\delta_{w}$ and $\delta_{h}$ are the structural variability of the waveguide at coordinates x,y, and the integral path C follows the layout of this arm. 

In order to calculate the distribution of the random phase error in a single waveguide or the random phase difference error between the random phase errors of the two arms of MZI, it is essential to properly model the underlying variability

\begin{itemize}
\item \textbf{Large wafer-to-wafer variability:} Across different wafer batches, the mean width and deposition thickness of waveguides exhibit significant variations, with a 3$\sigma$ range of approximately 5\% of the nominal linewidth.
\item \textbf{Small within-wafer variability:} Within a single wafer, the standard deviation of local width and thickness variations is relatively smaller, with a 3$\sigma$ range of approximately 5‰ of the nominal linewidth.
\item \textbf{Spatial continuity:} The variability has spatial continuity, meaning that gradual changes are more likely than abrupt fluctuations.
\end{itemize}

Therefore, a suitable variation model should satisfy the following three conditions:
\begin{itemize}
\item \textbf{Within-wafer local distribution:} For a specific wafer, the manufacturing deviation measured at an arbitrary point on any waveguide follows a distribution characterized by a mean of $\mu$ and a standard deviation of $\sigma_{\text{intra}}$.
\item \textbf{Wafer-to-wafer variation of mean values:} For any given wafer, the average width deviation of a standard waveguide follows a distribution with a mean of 0 and a standard deviation of $\sigma_{\text{inter}}$.
\item \textbf{Spatial continuity:} For any wafer, the covariance of variation between two arbitrarily selected points within the wafer depends exclusively on the Euclidean distance between them.
\end{itemize}

Given the aforementioned requirements, modeling the wafer-to-wafer standard deviation using a normal distribution and characterizing the within-wafer fabrication variability with a \textbf{two-dimensional stationary Gaussian process (GP)} is a reasonable approach.

Taking waveguide width variability as an example, we assume that due to fabrication imperfections, the center position of each waveguide remains fixed, while its width fluctuates. Let the width variation $\delta(x,y)$ be a stationary Gaussian process defined on the xOy plane:
\begin{equation}
\delta \left( x,y \right)\sim\mathcal{G}\mathcal{P}\left( m\left( x,y \right),\kappa\left( \mathbf{r},\mathbf{r}' \right) \right)
\end{equation}
where the mean function $m(x,y)$ is set to $\mu$, and the covariance function is defined as:
\begin{equation}
\kappa\left( \mathbf{r},\mathbf{r}' \right)=\sigma _{intra}^{2}{{e}^{-\frac{{{\left( x-x' \right)}^{2}}+{{\left( y-y' \right)}^{2}}}{r_{0}^{2}}}}\
\end{equation}

Here, $r_0$ represents the spatial correlation length, capturing the spatial continuity of fabrication variations.

\begin{figure}[hb]
\centering
\includegraphics[width=0.6\textwidth]{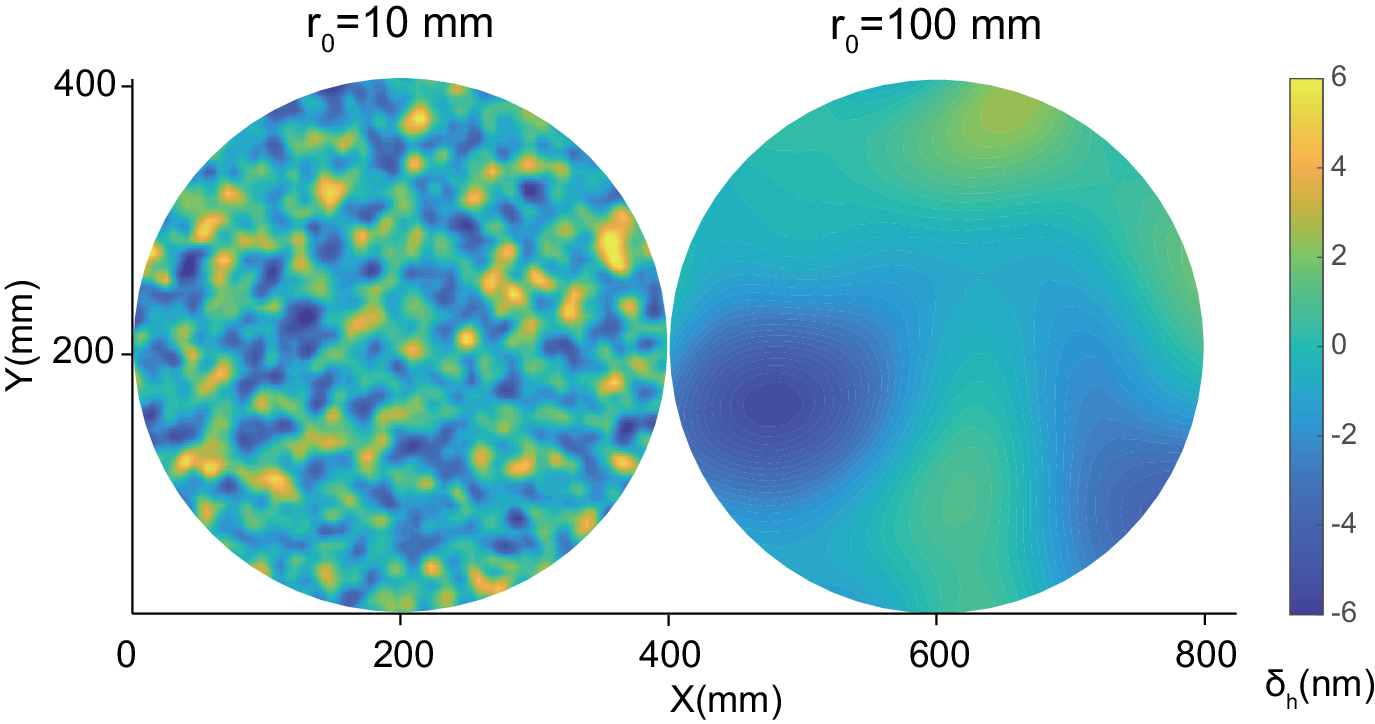}
\caption{The relationship of different spatial correlation lengths to the distribution of manufacturing variability}
\label{fig:wafer_thickness}
\end{figure}

Based on a stationary Gaussian distribution, we randomly generated wafer thickness maps with different spatial correlation lengths \( r_0 \). The results are shown in Fig.\ref{fig:wafer_thickness}, which intuitively illustrates the impact of the spatial correlation length on the distribution of manufacturing variability. As \( r_0 \) increases, adjacent points are more likely to exhibit similar manufacturing variations, reflecting a stronger spatial correlation.

By leveraging the stationary Gaussian process model, we can efficiently compute the random phase error distribution of any waveguide and further analyze the phase difference error or phase gradient error.

For the convenience of expression, we only consider the change of waveguide width. Consider an interferometric device controlled by \( N \) waveguides. The center trajectory of the \( i \)-th waveguide is denoted by \( C_i \), and its local effective index sensitivity to width variations is given by \(\xi_{wi}(x,y)\). Collecting the random phase errors from all \( N \) waveguides, we form the random phase  error vector ${{\mathbf{\Phi} }_{r}}={{\left[ \begin{matrix}   {{\Phi }_{1}} & {{\Phi }_{2}} & \cdots  & {{\Phi }_{N}}  \\\end{matrix} \right]}^{T}}$
 .Under our previous assumptions, the joint distribution of these N random phases follows a multivariate normal distribution $\mathbf{\Phi} \sim \mathcal{N}(\mathbf{\mu}, \mathbf{\Sigma})$. The mean and covariance matrices follow equations \ref{def:mu_multi} and \ref{eq:CovMatrix}, respectively

\begin{equation}
    {{\mu }_{i}}={{k}_{0}}\mu \int_{{{C}_{i}}}{{{\xi }_{wi}}}(x,y)ds
    \label{def:mu_multi}
\end{equation}

\begin{equation}
    {{\Sigma }_{ij}}=k_{0}^{2}\int\limits_{{{C}_{i}}}{\int\limits_{{{C}_{j}}}{{{\xi }_{wi}}\left( \mathbf{r} \right){{\xi }_{wj}}\left( \mathbf{r}' \right)\kappa\left( \mathbf{r},\mathbf{r}' \right)ds'ds}}
    \label{eq:CovMatrix}
\end{equation}

After obtaining the covariance matrix $\mathbf{\sigma}$by means of numerical integration or analytic method, we can analyze the random phase difference error of two particular waveguides or the random phase gradient error of this group of waveguides.

\section{Example}\label{sec3}
Although numerical methods are generally required to analyze realistic structures, the theoretical framework described above allows us to derive analytical expressions for the random phase error distributions in some simple cases. In this section, we analyze two basic structures, parallel array of straight waveguides and concentric array of curved waveguides and compute their random phase error distributions using the proposed theory. This analysis not only illustrates how the cumulative effects of spatial correlation on random errors but also provides useful guidance for the design of practical phase shifter configurations.

\subsection{Parallel array of straight waveguides}

In many photonic devices, such as Mach–Zehnder interferometers (MZI) and arrayed waveguide gratings (AWG), straight waveguides are arranged in parallel. Although there is no energy coupling between them, they provide a phase gradient. Because the center-to-center spacing of these waveguides is typically smaller than the spatial correlation length $r_0$, the random phase errors in this set of waveguides are correlated rather than independent. 

 In this section, we first compute the mean and covariance of the random phase errors for N parallel straight waveguides. We then analyze two special cases, and finally introduce a method for analyzing phase gradient variations.

 We assume that  N straight waveguides are arranged along the x-axis, all starting at the same x-coordinate but having different lengths $L_i$. Their positions along the y-axis are given by $y_i$. Each waveguide exhibits a constant effective refractive index sensitivity, denoted by $\xi_i$. The spatial correlation length is $r_w$, and the standard deviation within each waveguide is $\sigma_{\text{w}\_\text{intra}}$ . Using equations \ref{def:mu_multi} and \ref{eq:CovMatrix}, we calculate the mean and covariance matrix.

\begin{equation}
    {{\mu }_{i}}={{k}_{0}}\mu {{\xi }_{i}}{{L}_{i}}
    \label{eq:mean_wgpair}
\end{equation}
\begin{equation}
{{\Sigma }_{ij}}=k_{0}^{2}{{\xi }_{i}}{{\xi }_{j}}\sigma _{w\_\text{intra}}^{2}L_{\text{meff}}^2
    \label{eq:covmatrix_wgpair}
\end{equation}
Where effective length $L_{\text{eff}}$ represents the effective cumulative length of random error under the influence of spatial correlation length,  and mutual effective length $L_{\text{meff}}$  represents the interaction of the cumulative effect of random errors between two waveguides due to spatial correlation. They are defined as:
\begin{equation}
   L_{\text{meff}}^2= {{e}^{-\frac{{{\left(y_{i}-y_{j}\right)}^{2}}}{r_{w}^{2}}}}\frac{L_{\text{eff}}^{2}\left( {{L}_{i}},{{r}_{w}} \right)+L_{\text{eff}}^{2}\left( {{L}_{j}},{{r}_{w}} \right)-L_{\text{eff}}^{2}\left( {{L}_{i}}-{{L}_{j}},{{r}_{w}} \right)}{2}
    \label{def:Ceff_for_psw}
\end{equation}
\begin{equation}
{{L}_{\text{eff}}^2}\left(L,r\right)=\sqrt{\pi}L{r}\cdot \text{erf}\left( \frac{L}{r} \right)-r^{2}\left( 1-{{e}^{-{{\left( \frac{L}{{r}} \right)}^{2}}}} \right)
\label{def:Leff}
\end{equation}
Where, the effective length can be simplified when the waveguide length is much less than or much larger than the spatially relevant length:

\begin{equation}
L_{\text{eff}}^{2}(L, r) =
\begin{cases}
  L^2, & L \ll r \\
  \sqrt{\pi} L r, & L \gg r
\end{cases}
\end{equation}

\begin{figure}[hb]
\centering
\includegraphics[width=0.6\textwidth]{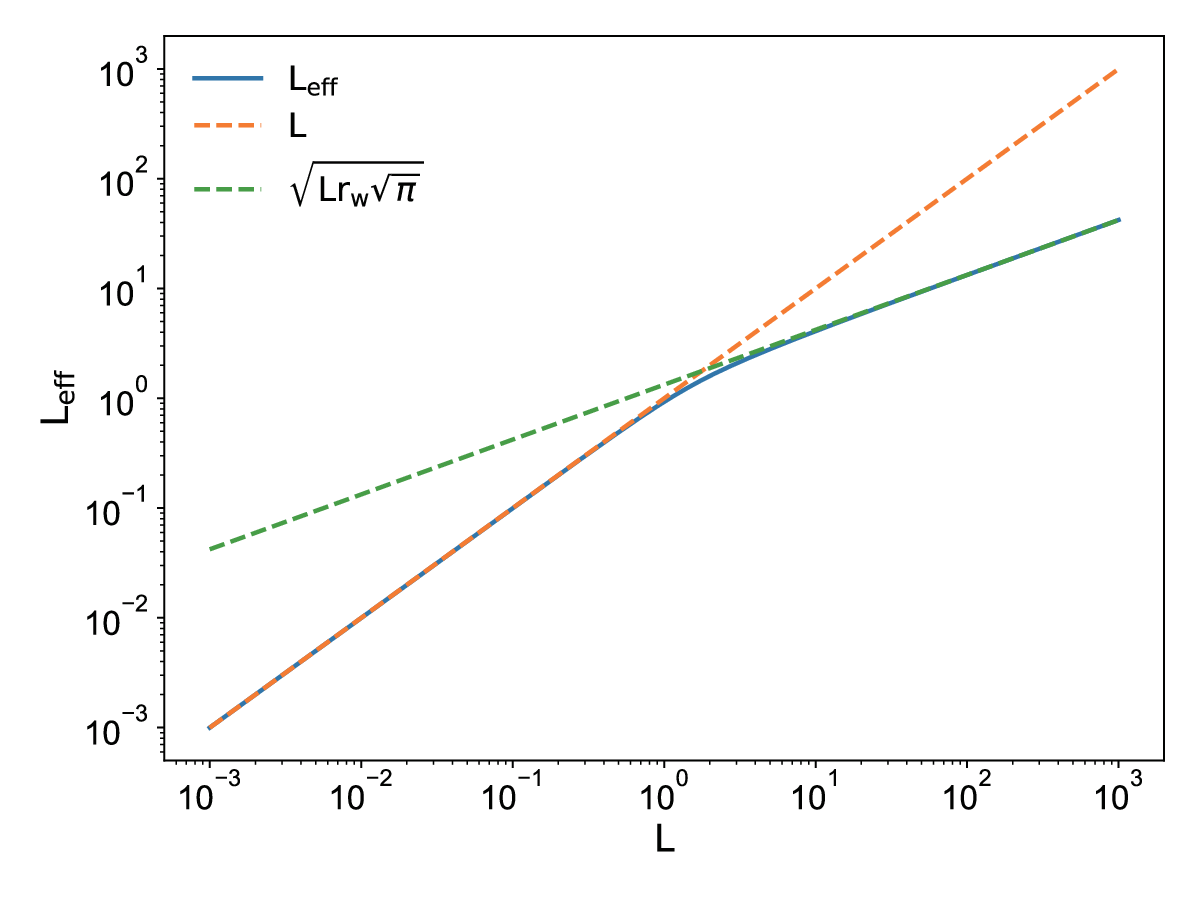}
\caption{Effective length $L_{\text{eff}}$  as a function of length $L$ in Straight waveguide ($r_{w}$=1)}
\label{fig:function_Leff}
\end{figure}

The function image of $L_{\text{eff}}$ is shown in the Fig. \ref{fig:function_Leff}. This result indicates that, for waveguide lengths much shorter than the spatial correlation length, the effective error accumulation becomes equivalent to the physical length. Conversely, when the waveguide is significantly longer than the correlation length, the effective length becomes proportional to the square root of the physical length. This reflects a saturation effect in error accumulation due to the finite spatial correlation length.

Using the mean and covariance matrices, we can directly calculate quantities of interest. Taking the calculation of the mean and variance of the random phase difference error between waveguide 1 and waveguide 2 as an example, the statistic we need to calculate is actually  $\Delta\Phi_r=\Phi_1-\Phi_2=\mathbf{A}\mathbf{\Phi}$, where $\mathbf{A}=[1,-1]^T$. The specific calculation is as follows:
\begin{equation}
\mathbb{E}\left(\Delta\Phi_r\right)=\mathbf{A}\mathbb{E}\left({\mathbf{\Phi}}\right)=k_0\mu\left(\xi_{1}L_{1}-\xi_{2}L_{2}\right)
\end{equation}
\begin{equation}
    \text{Var}\left( \Delta {{\Phi }_{r}} \right)={{\mathbf{A}}^{T}}\mathbf{\Sigma A}={{\Sigma }_{11}}+{{\Sigma }_{22}}-{{\Sigma }_{12}}-{{\Sigma }_{21}}
\end{equation}
When $L_1\ll L$ and $L_2\ll L$, the variance is:
\begin{equation}
\text{Var}\left( \Delta {{\Phi }_{r}} \right)=k_{0}^{2}\sigma _{\text{w}\_\text{intra}}^{2}\left[ {{\xi }_{1}}^{2}L_{1}^{2}+\xi _{2}^{2}L_{2}^{2}-2{{\xi }_{1}}{{\xi }_{2}}{{L}_{1}}{{L}_{2}}{{e}^{-\frac{{{\left( {{y}_{1}}-{{y}_{2}} \right)}^{2}}}{r_{w}^{2}}}} \right]
\end{equation}
We can minimize the mean and variance of the random phase difference error by taking  $\xi_{1}L_1=\xi_{2}L_2$ in the same time,  the result is:
\begin{equation}
\mathbb{E}\left(\Delta\Phi_r\right)=0
\end{equation}
\begin{equation}
\text{Var}\left( \Delta {{\Phi }_{r}} \right)=2k_{0}^{2}\sigma _{\text{w}\_\text{intra}}^{2}{{\xi }_{1}}{{\xi }_{2}}{{L}_{1}}{{L}_{2}}\left[ 1-{{e}^{-\frac{{{\left( {{y}_{1}}-{{y}_{2}} \right)}^{2}}}{r_{w}^{2}}}} \right]
\end{equation}

When $L_1\gg L$ and $L_2\gg L$, the variance is:
\begin{equation}
    \text{Var}\left( \Delta {{\Phi }_{r}} \right)=k_{0}^{2}\sigma _{\text{w}\_\text{intra}}^{2}\sqrt{\pi }r\left[ {{\xi }_{1}}^{2}{{L}_{1}}+\xi _{2}^{2}{{L}_{2}}-2{{\xi }_{1}}{{\xi }_{2}}\sqrt{{{L}_{1}}{{L}_{2}}}{{e}^{-\frac{{{\left( {{y}_{1}}-{{y}_{2}} \right)}^{2}}}{r_{w}^{2}}}} \right]
\end{equation}
We can not minimize the mean and variance of the random phase difference error in the same time, unless we choose $L_1=L_2=L$ and  $\xi_{1}=\xi_{2}=\xi$, the result is:
\begin{equation}
    \mathbb{E}\left(\Delta\Phi_r\right)=0
    \label{eq:DeltaPhi_mean_wgpair}
\end{equation}
\begin{equation}
\text{Var}\left( \Delta {{\Phi }_{r}} \right)=2\sqrt{\pi }k_{0}^{2}\sigma _{\text{w}\_\text{intra}}^{2}rL{{\xi }^{2}}\left[ 1-{{e}^{-\frac{{{\left( {{y}_{1}}-{{y}_{2}} \right)}^{2}}}{r_{w}^{2}}}} \right]
\end{equation}

This means that for a pair of adjacent waveguides, the expected value of their random phase difference error is equal to the expected phase difference of a single waveguide. The variance of the phase difference depends on the effective length of the waveguides and their center spacing. It is worth noting that when the length and sensitivity of the two waveguides are the same, the desired phase difference becomes zero. With the increase of the spacing, the phase difference variance increases and tends to saturation, indicating that the spatial correlation is weakened and the phase error is statistically independent. When waveguides are placed close to each other, tolerance optimization of the random phase errors can be achieved not only by widening the waveguides or reducing the index sensitivity $\xi$, but also by ensuring that adjacent waveguides share the same $\xi$. 

In the case of AWG, we are concerned with the phase gradient. Ideally, if the manufacturing error causes each waveguide to have exactly the same random phase error, there will be no change in device performance.  We can calculate a random phase error vector \textbf{$\left[\Phi_1;\Phi_2;...;\Phi_N\right]$ }with a projection on vector $I=\left[1;1;1;...1\right]/\sqrt{N}$ to evaluate the effect of AWG performance degradation. In this case, the length of the projection is:
\begin{equation}
    {{I}_{proj}}=\frac{\mathbf{\Phi }\cdot \mathbf{I}}{\left\| \mathbf{I} \right\|}=\mathbf{\Phi }\cdot \mathbf{I}
\end{equation}
The mean and variance of this projection length are:
\begin{equation}
    \mathbb{E}\left( {{I}_{proj}} \right)=\mathbf{I}\cdot \mathbb{E}\left( {{\Phi }_{i}} \right)
\end{equation}
\begin{equation}
    \text{Var}\left( {{I}_{proj}} \right)={{\mathbf{I}}^{T}}\mathbf{\Sigma I}
\end{equation}
When we choose the absolute value of the cosine of the included Angle as the statistic to measure the goodness of the AWG random phase gradient error, the reasoning becomes extremely complicated, and this paper fails to solve this problem.

\subsection{Concentric Array of Curved waveguides}
Curved waveguides are essential components in integrated photonic circuits. Compared to straight waveguides of the same arc length, they typically exhibit different random phase errors due to their curved geometry and distributed path. In this subsection, we analyze the behavior of random phase errors among a group of concentric circular waveguides. Each waveguide has a distinct radius $R_i$ and a unique effective refractive index sensitivity $\xi_i$, but they all share the same arc angle. Based on the proposed theoretical framwork, the mean vector and covariance matrix of the random phase errors distribution for this structure can be calculated as follows:

\begin{equation}
    \mathbb{E}\left(\Phi_{ri}\right)=k_0\xi_{i}\mu_{w}R_{i}\theta_0
    \label{eq:Phi_mean_curvewg}
\end{equation}
\begin{equation}
\Sigma_{ij}=k_0^2\xi_{i}\xi_{j}\sigma_{\text{w}\_\text{intra}}^2L_{\text{meff}}^2\left(R_i,R_j,\theta_0,r\right)
    \label{eq:Phi_var_curvewg}
\end{equation}
Where mutual effective arc length $L_{\text{meff}}$ is defined as:

\begin{equation}
L_{\text{meff}}^{2}\left( {{R}_{i}},{{R}_{j}},{{\theta }_{0}},r \right)=2{{R}_{i}}{{R}_{j}}{{e}^{-\frac{R_{i}^{2}+R_{j}^{2}}{{{r}^{2}}}}}\int_{0}^{{{\theta }_{0}}}{\left( {{\theta }_{0}}-u \right){{e}^{\frac{2{{R}_{i}}{{R}_{j}}\cos u}{{{r}^{2}}}}}du}
    \label{eq:Curvewgarray_Sigma}
\end{equation}
We can give approximate results in two limit cases by:
\begin{equation}
L_{\text{meff}}^{2}\left( R_{i}, R_{j}, \theta_{0}, r \right) =
\begin{cases}
  R_{i} R_{j} \theta_{0}^{2}e^{-\frac{\left(R_i-R_j\right)^2}{r^2}}, & R_{i} R_{j} \ll r^{2} \\
  \theta_{0} r \sqrt{\pi R_{i} R_{j}}e^{-\frac{\left(R_i-R_j\right)^2}{r^2}}, & R_{i} R_{j} \gg r^{2}
\end{cases}
\label{eq:Ceff_simplify}
\end{equation}
Taking the calculation of the mean and variance of the random phase difference error between curved waveguide 1 and curved waveguide 2 as an example, the statistic we need to calculate is also  $\Delta\Phi_r=\Phi_1-\Phi_2=\mathbf{A}\mathbf{\Phi}$, where $\mathbf{A}=[1,-1]^T$. The specific calculation is as follows:

\begin{equation}
\mathbb{E}\left(\Delta\Phi_r\right)=\mathbf{A}\mathbb{E}\left({\mathbf{\Phi}}\right)=k_0\mu_w\left(\xi_{1}R_{1}-\xi_{2}R_{2}\right)\theta_0
\end{equation}
\begin{equation}
    \text{Var}\left( \Delta {{\Phi }_{r}} \right)={{\mathbf{A}}^{T}}\mathbf{\Sigma A}={{\Sigma }_{11}}+{{\Sigma }_{22}}-{{\Sigma }_{12}}-{{\Sigma }_{21}}
\end{equation}
When $ R_{i} R_{j} \ll r^{2}$, the variance is:
\begin{equation}
\text{Var}\left( \Delta {{\Phi }_{r}} \right)=k_{0}^{2}\sigma _{\text{w}\_\text{intra}}^{2}\theta _{0}^{2}\left\{ {{\left( {{\xi }_{1}}{{R}_{1}}-{{\xi }_{2}}{{R}_{2}} \right)}^{2}}+2{{\xi }_{1}}{{\xi }_{2}}{{R}_{1}}{{R}_{2}}\left[ 1-{{e}^{-\frac{{{\left( {{R}_{1}}-{{R}_{2}} \right)}^{2}}}{r_{w}^{2}}}} \right] \right\}
\end{equation}
We can minimize the mean and variance of the random phase difference error by taking  $\xi_{1}R_1=\xi_{2}R_2$ in the same time,  the result is:
\begin{equation}
\mathbb{E}\left(\Delta\Phi_r\right)=0
\end{equation}
\begin{equation}
\text{Var}\left( \Delta {{\Phi }_{r}} \right)=2k_{0}^{2}\sigma _{\text{w}\_\text{intra}}^{2}\theta _{0}^{2}{{\xi }_{1}}{{\xi }_{2}}{{R}_{1}}{{R}_{2}}\left[ 1-{{e}^{-\frac{{{\left( {{R}_{1}}-{{R}_{2}} \right)}^{2}}}{r_{w}^{2}}}} \right]
\end{equation}
However, when $ R_{i} R_{j} \gg r^{2}$, we can not minimize the mean and variance in the same time. It is different from parallel array of straight waveguides model. But controling  $\xi_{1}R_1=\xi_{2}R_2$ is always a good ideal.
\section{Application}\label{sec4}
\subsection{Calibration-Free Integrated Magneto Circulator on SiN}
Based on the above theoretical framework and methodology, we first conduct a theoretical investigation of a thermally calibration-free MZI-based integrated magneto-optical circulator. The circulator consists of magneto-optical waveguides that provide nonreciprocal phase shifts and conventional waveguides that provide reciprocal phase shifts, as illustrated in the figure. Fabrication imperfections introduce random phase difference error between the two arms, which in turn lead to a shift in the central operating wavelength.

\begin{figure}[h]
\centering
\includegraphics[width=0.6\textwidth]{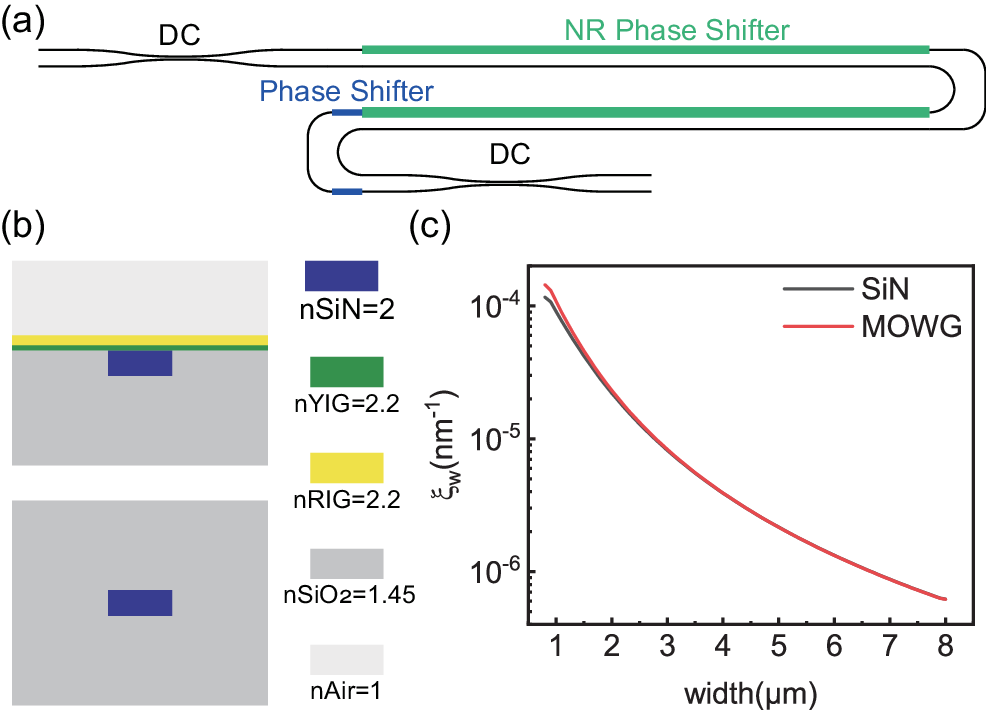}
\caption{(a)Schematic of SiN integrated magneto-optical Circulator (b)Structure diagram of magneto-optical SiN waveguide and SiN waveguide (c)Sensitivity of TM0 mode effective refractive index vs SiN width }
\label{fig:Struct}
\end{figure}

Fig. \ref{fig:Struct}(a) illustrates the basic structure of the integrated magneto-optical circulator. Two nonreciprocal phase-shifting magneto-optical waveguides are arranged in a push-pull configuration, and the use of folded waveguides enables the device to operate under a unidirectional magnetization state. Fig. \ref{fig:Struct}(b) shows the cross-sectional structures of the SiN magneto-optical waveguide and the standard SiN waveguide, along with the selected refractive indices of the materials. The SiN thickness is fixed at 400 nm, the YIG seed layer thickness is 50 nm with a Verdet constant of 150 deg/cm, and the RIG film thickness is 120 nm with a Verdet constant of 4000 deg/cm.Fig. \ref{fig:Struct}(c) presents the effective refractive index sensitivity of the TM\textsubscript{0} mode in the magneto-optical waveguide as a function of the SiN waveguide width.

Due to the broken vertical symmetry introduced by the magneto-optical layers, increasing the width of the magneto-optical waveguide not only results in multimode behavior of the TM\textsubscript{0} mode, but may also lead to the excitation of odd-symmetry modes such as TE\textsubscript{2n+1}. Therefore, instead of reducing the index sensitivity $\xi$ by widening the waveguide, we exploit the spatial correlation of fabrication variations to optimize the random phase difference error.

\begin{figure}[h]
\centering
\includegraphics[width=0.8\textwidth]{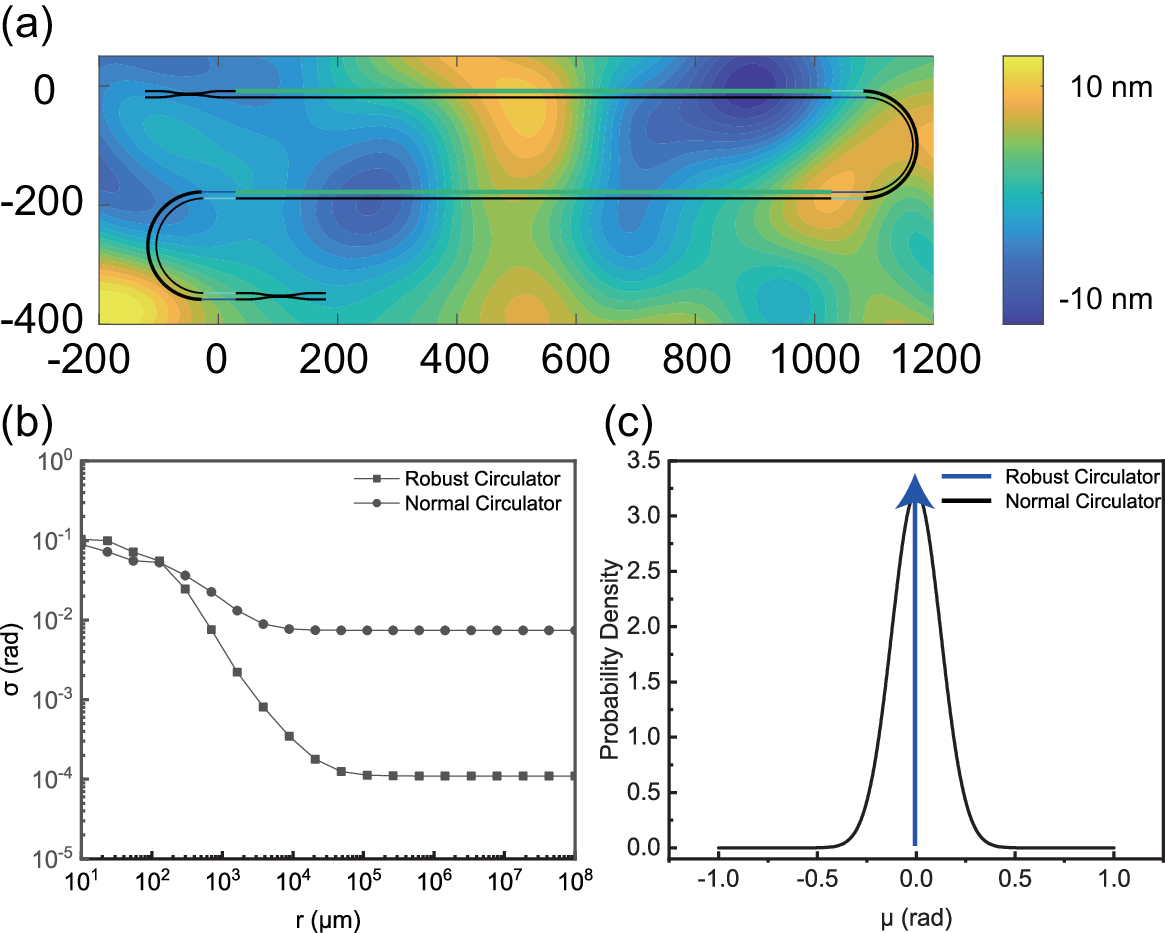}
\caption{(a)Schematic of Calibration-free SiN integrated magneto-optical Circulator (b) The relation between the standard deviation of the random phase difference and the spatial correlation length $r_0$ (c)Distribution of the mean of the random phase difference (select $3\sigma_{\textbf{inter}}$=50 nm) }
\label{fig:Calibration-free CR}
\end{figure}

For the integrated magneto-optical circulator structure, the required length of the magneto-optical waveguide is approximately 943 \textmu m, and the bending radius of the folded waveguides should be larger than 75 \textmu m. According to the estimation based on data from Ref.\cite{bogaerts_predicting_2019}, the spatial correlation length of width variations is approximately from 161 \textmu m to 500 \textmu m. Using numerical integration, Eq.\ref{def:mu_multi} and Eq.\ref{eq:CovMatrix} , we compute the mean and covariance matrix of the random phase error in both arms. The calculation results are shown in Fig. \ref{fig:Calibration-free CR}.

Fig. \ref{fig:Calibration-free CR}(a) illustrates the design concept of the calibration-free integrated magneto-optical circulator. Our design leverages the conclusions from previous analyses: for the magneto-optical waveguide (represented by the green line segment) and the standard waveguide pair, the SiN core widths are selected to ensure that both exhibit the same index sensitivity \( \xi \); for the bent waveguides, the condition \( R_{\text{inner}} \xi_{\text{inner}} = R_{\text{outer}} \xi_{\text{outer}} \) is maintained; and for the reciprocal phase-shifting segments (indicated by the blue and light blue line segments), the products \( L_{\text{rps1}} \xi_{1} = L_{\text{rps2}} \xi_{2} \)  are controlled, with the lengths chosen appropriately to achieve a free spectral range of 100~nm. Fig. \ref{fig:Calibration-free CR} (b) and (c) display the relationship between the standard deviation of the random phase difference error in the integrated calibration-free magneto-optical circulator and the spatial correlation length \( r \) across a single wafer. Compared to a circulator without calibration-free design optimization, the optimized design achieves a lower standard deviation for larger spatial correlation lengths. Additionally, the optimized circulator maintains a mean phase difference of zero, thus remaining unaffected by average width variations across wafers.
\section{Conclusion}\label{sec5}
As the scale of integrated photonic chips continues to increase, post-layout-level performance analysis and yield prediction are becoming critically important. In this work, we propose a theoretical framework for modeling and analyzing random phase errors that avoids reliance on computationally intensive Monte Carlo simulations, achieving high computational efficiency without compromising accuracy. The framework is highly generalizable and can be seamlessly integrated into existing simulation workflows and EPDA tools.

Furthermore, by combining our method with the compact modeling approach introduced in Ref.~\cite{zhang_compact_2025} and the theory of stochastic differential equations, it may be extended to analyze other types of photonic devices, such as adiabatic couplers and multimode interference (MMI) structures. The proposed framework also allows for flexible adaptation to different manufacturing variability profiles through modification of the covariance function.

We believe that combining this approach with commercial photonic link simulation tools holds great potential for improving design-for-manufacturability and reliability in large-scale photonic integrated circuits.

\backmatter

\section*{Declarations}
The English translation of this paper was assisted by AI generation and has been carefully reviewed by the author. However, any inaccuracies or misinterpretations remain the responsibility of the author.

\subsection*{Conflict of interest/Competing interests}
The author declares no conflict of interest.

\subsection*{Data availability}
The datasets generated and analyzed during the current study are available from the corresponding author upon reasonable request.

\subsection*{Code availability}
The code used for this study is available from the corresponding author upon reasonable request.

\bigskip

\begin{appendices}
\end{appendices}


\bibliography{sn-bibliography}

\end{document}